\documentstyle[11pt,newpasp,twoside,epsf]{article}
\def\edcomment#1{\iffalse\marginpar{\raggedright\sl#1\/}\else\relax\fi}
\reversemarginpar

\begin{document}

\title{
The Extragalactic Ionizing Background at Low Redshift}

\author{Jennifer Scott, Jill Bechtold, Miwa Morita}
\affil{Steward Observatory, University of Arizona, Tucson, AZ 85721}

\author{Adam Dobrzycki}
\affil{Harvard-Smithsonian Center for Astrophysics, 60 Garden Street,
Cambridge, MA 02138}

\author{Varsha P. Kulkarni}
\affil{University of South Carolina, Department of Physics and Astronomy, 
Columbia, SC  29208}

\begin{abstract}
We present a measurement of the mean intensity of the 
hydrogen-ionizing background radiation field at low redshift using 
906 Ly$\alpha$ absorption lines in 151 quasar spectra
from the archives of the Faint Object Spectrograph (FOS) on the
Hubble Space Telescope (HST). 
Using a maximum likelihood technique and
the best estimates possible for each QSO's Lyman limit flux and
systemic redshift,  we find
$J(\nu_{0})$= 7.6$^{+9.4}_{-3.0} \times 10^{-23}$ ergs s$^{-1}$ cm$^{-2}$ Hz$^{-1}$ sr$^{-1}$
at $0.03 < z < 1.67$.  This is in good agreement with the mean intensity expected
from models of the background which incorporate only the known quasar population.
When the
sample is divided into two subsamples, consisting of lines with
$z < 1$ and $z > 1$,  the values of $J(\nu_{0})$ found are
6.5$^{+38.}_{-1.6} \times 10^{-23}$ ergs s$^{-1}$ cm$^{-2}$ Hz$^{-1}$ sr$^{-1}$,
and
1.0$^{+3.8}_{-0.2} \times 10^{-22}$ ergs s$^{-1}$ cm$^{-2}$ Hz$^{-1}$ sr$^{-1}$,
respectively, indicating that the mean intensity of the
background is evolving over the redshift range of this data set.
Relaxing the assumption that the spectral shapes of the sample spectra and the
background are identical, the best fit HI photoionization rates are found to be
$6.7 \times 10^{-13}$ s$^{-1}$ for all redshifts, and $1.9 \times 10^{-13}$ s$^{-1}$
and $1.3 \times 10^{-12}$ s$^{-1}$ for $z < 1$ and $z > 1$, respectively.
\end{abstract}

\section{Mean intensity of the UV Background}
The maximum likelihood method for measuring $J(\nu_{0})$ as presented by
Kulkarni \& Fall (1993, KF93) consists of constructing a
likelihood function of the form
\begin{equation}
L= \prod_{a}{\it f}(N_{a},z_{a}) \prod_{Q} {\rm exp}[ - \int^{z^{Q}_{max}}
_{z^{Q}_{min}} dz \int^{\infty}_{N^{Q}_{min}} {\it f}(N,z) dN ] \label{eq:maxlike},
\end{equation}
where
\begin{equation}
{\it f}(N,z) = AN^{-\beta}(1+z)^{\gamma}[1+\omega(z)]^{-(\beta-1)}.
\label{eq:fnz}
\end{equation}
Using the values of $\gamma$ and $A_{0}$ from a separate maximum likelihood
analysis, and a value of $\beta$ from studies with high resolution data, eg.
$\beta=1.46$ from Hu et al.\ (1995), the search for the best-fit value of
$J(\nu_{0})$ consists of finding the value that maximizes this function,
fixing the other parameters.

\begin{figure}
\plotfiddle{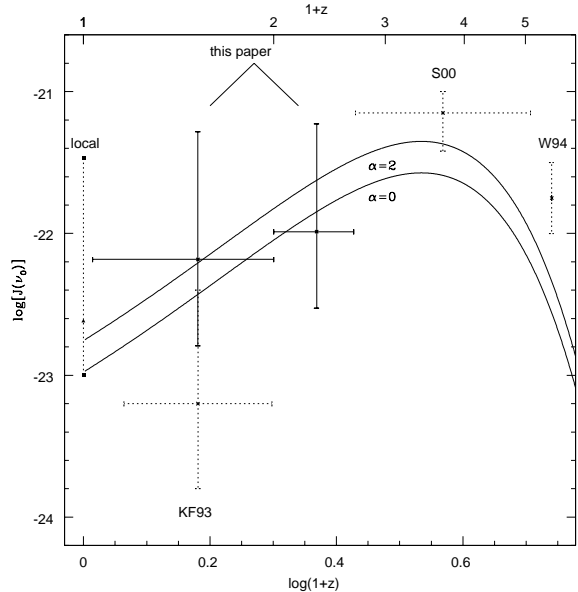}{2.75in}{0}{40}{40}{-130}{-75}
\caption{
log[$J(\nu_{0})$] versus redshift:
(filled triangle)- Shull et al.\ (1999);
(filled squares \& dotted line)- limits from H$\alpha$ imaging;
(crosses)- our results for $z < 1$ and $z > 1$;
points with dotted error bars centered at $z\sim$0.6,3.,4.5 are results from
KF93, Paper II, and Williger et al.\ (1994), respectively;
(solid curves)- HM96 models for two values of the global source spectral
index, $\alpha$
\label{fig:allzcomp} }
\end{figure}

The models
of Haardt \& Madau (1996, HM96) predict that the UV background arising
from QSOs drops by over an order of magnitude from $z=2.5$ to $z=0$.
We therefore
divide the sample into low and high redshift subsamples at $z=1$ and 
solve for $J(\nu_{0})$.
These results, listed in Table 1.
confirm some evolution in $J(\nu_{0})$, though not at a high level of significance.
The maximum likelihood analysis yields
log[$J(\nu_{0})$] = -22.18$^{+0.90}_{-0.61}$ at $z < 1$ and
log[$J(\nu_{0})$] = -21.98$^{+0.76}_{-0.54}$ at $z > 1$.
These results are also shown in Figure 1.

\begin{table}
\caption{Measurements of $J(\nu_{0})$}
\begin{tabular}{llcccccl}
\tableline
Sample\tablenotemark{a}  &${\cal N}_{lines}\tablenotemark{b}$
&$\gamma$,$A$ &$\beta$ &b &log[$J(\nu_{0})$] &Q$_{\rm KS}$\tablenotemark{c}  \\
\tableline
1\dotfill  &259  &0.8298, 6.73524 &1.46 &35 &-22.11$^{+0.51}_{-0.40}$ &0.80 \\
1a\dotfill &162  &1.5082, 4.92095 &1.46 &35 &-22.18$^{+0.90}_{-0.61}$ &0.64 \\
1b\dotfill &97   &-0.8702,26.1886 &1.46 &35 &-21.98$^{+0.76}_{-0.54}$ &0.98 \\
2\dotfill  &289  &0.1502, 12.0134 &1.46 &35 &-22.03$^{+0.44}_{-0.37}$ &0.30 \\
\tableline
\tableline
\tablenotetext{a}{Sample number- (1) All lines with W$>$0.32 $\AA$,
(1a) z$ < $1, (1b) z$ > $1;
(2) All lines with W$ > $0.24 $\AA$}
\tablenotetext{b}{Number of Ly$\alpha$ forest lines in sample}
\tablenotetext{c}{K-S probability}
\end{tabular}
\end{table}

Including associated absorbers, damped Ly$\alpha$ absorbers, or blazars
in the proximity effect analysis has little effect on
the results.  One might expect associated absorbers to
reduce the magnitude of the observed proximity effect and hence
cause $J(\nu_{0})$ to be overestimated.  The value found including
the 45 associated absorbers in our sample is indeed larger,
log[$J(\nu_{0})$]=-21.74$^{+0.55}_{-0.39}$, versus
log[$J(\nu_{0})$]=-22.11$^{+0.52}_{-0.40}$, but not significantly so.
Likewise, if the intervening dust extinction in damped Ly$\alpha$ absorbers
is significant, including these objects in our analysis could 
cause us to overestimate the magnitude of the proximity effect and
hence underestimate $J(\nu_{0})$.  However, the inclusion of these 7 
objects only negligibly reduces the value of $J(\nu_{0})$ derived.
QSO variability on timescales less than $\sim$10$^{5}$ years would be expected
to smooth out the proximity effect distribution (Bajtlik, Duncan, \& Ostriker 1988).  
However, the inclusion
of 6 blazars in the sample, all at $z < 1$, resulted in no discernible change in $J(\nu_{0})$.
The sample used in the analysis of HI ionization rates discussed below
includes all of these objects.

For each solution, we also execute a Kolmogorov-Smirnov (KS) test.
The KS test provides a measure of how well the assumed parent
distribution of lines with respect to redshift,
given by Equation 2,
reflects the true redshift distribution of lines.
The KS probability, Q$_{\rm KS}$, indicates the probability that
a value of the KS statistic larger than the one calculated could have
occurred by chance if the assumed parent is correct.
The KS probability associated with each solution for $J(\nu_{0})$
is listed in column 7 of Table 1.

\section{HI Ionization Rate}
\label{sec-gam}

Solving for the HI ionization rate,
\begin{equation}
\Gamma=\int_{\nu_{0}}^{\infty} \frac{4 \pi J(\nu) \sigma_{HI}(\nu)}{h \nu} d\nu
\; \; {\rm s^{-1}},
\label{eq:gamma}
\end{equation}
instead of for $J(\nu_{0})$ avoids the
assumption that the spectral indicies of the QSO and
the background are identical.  We modified our maximum likelihood
code to conduct the search for this quantity and the
results are listed in Table 2.
Evolution in the UV background is more apparent in the HI
ionization rate than in the solutions for $J(\nu_{0})$.
The result at $z > 1$ is 6.5 times larger than  that at $z < 1$.

\begin{table}
\caption{HI Ionization Rates}
\begin{tabular}{lcccc}
\tableline
Sample\tablenotemark{a}  &$\gamma$,$A$ &$\beta$ &b &log[$\Gamma_{\rm HI}$] \\
\tableline
1\dotfill  &0.6925,7.64986 &1.46 &35 &-12.17$^{+0.50}_{-0.40}$ \\
1a\dotfill &0.8452,7.10998 &1.46 &35 &-12.70$^{+0.74}_{-0.51}$ \\
1b\dotfill &0.7209,7.29421 &1.46 &35 &-11.88$^{+0.74}_{-0.50}$ \\
1\dotfill  &0.6925,7.20637 &1.46 &35 &-12.67,1.73\tablenotemark{1} \\
\tableline
\tableline
\tablenotetext{a}{(1) All lines with W$>$0.32 $\AA$, (1a) z$ < $1, (1b) z$ > $1}
\tablenotetext{1}{Maximum Likelihood solution for A,B
(see \S 2, Equ.\ 4)}
\end{tabular}
\end{table}

We also parametrize the evolution of the HI ionization rate
as a power law:
\begin{equation}
\Gamma(z)= A(1+z)^{B}
\label{equ:plgam}
\end{equation}
and solve for the parameters A and B in both the constant and variable threshold
cases.  The values we find are
also listed in Table 2 and plotted in Figure 2.

\begin{figure}
\plotfiddle{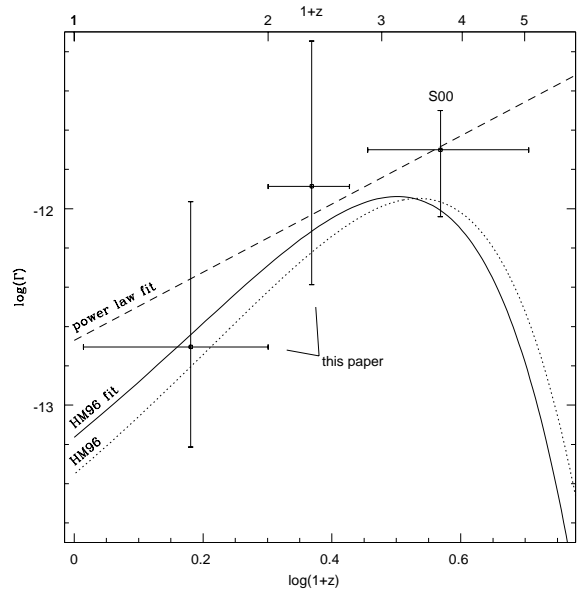}{2.75in}{0}{40}{40}{-130}{-75}
\caption{HI ionization rate versus redshift:
(points)- constant equivalent width threshold maximum likelihood solutions
from this paper, $z < 1$ and $z > 1$ and from Paper II for
$1.7 < z < 3.8$;
(dashed line)- Solution to Equ.\ 4
for HST/FOS data alone;
(solid line)- Solution to  Equ.\ 5
for HST/FOS data combined with high redshift data from Paper II,
(dotted line)- HM96 solution to Equ.\ 5
\label{fig:gam} }
\end{figure}

HM96 parametrize their models of the HI ionization rate as a function
of redshift:
\begin{equation}
\Gamma(z)= A(1+z)^{B} \exp \left( \frac{-(z-z_{c})^{2}}{S} \right)
\label{equ:hmgam}
\end{equation}
We combine our data set with that of Scott et al.\ 2000b (Paper II) to solve for the
parameters $A$, $B$, $zc$, and $S$.  We find $(A,B,z_{c},S)$=($7.6 \times 10^{-13}$,0.35,2.07,1.77),
while the HM96
parameters for $q_{0}=0.5$ are ($6.7 \times 10^{-13}$,0.43,2.30,1.95).
These results are shown in  Figure 2, and
the HM96 parametrization is also shown for comparison.

\section{Ly$\alpha$ forest line density}
\label{sec-dndz}

The number density evolution of Ly$\alpha$ absorbers over the
redshift range $z=0-5$ cannot be approximated with a single power
law. There is a significant break in the slope of the
line number density with respect to redshift, near $z\approx1.7$.
Dav\'{e} et al.\ (1999) show from hydrodynamical simulations of the low
redshift Lyman $\alpha$ forest, that the evolution of the line density
is sensitive mainly to the HI photoionization rate, but also to the  
evolution of structure.  The flattening of $d{\cal N}/dz$ observed by
Weymann et al.\ (1998) is mostly attributed to a dramatic decline in
$\Gamma(z)$ with decreasing $z$.
Dav\'{e} et al.\ (1999) derive an expression for the density of Lyman $\alpha$ forest
lines per unit redshift as a function of the HI photoionization rate:
\begin{equation}
\frac{d{\cal N}}{dz} = C [ (1+z)^{5} \Gamma^{-1}_{\rm HI}(z)]^{\beta-1} H^{-1}(z),
\label{equ:dave}
\end{equation}
where $C$ is the normalization at some fiducial redshift which we choose to be $z=0$
and $\Gamma (z)$ can be expressed by Equ.\ 5.

We fit the FOS and MMT absorption line data,
binned in $d{\cal N}/dz$, presented in Dobrzycki et al.\ (2001, Paper IV) and in
Scott et al.\ (2000a, Paper I), to this function
in order to derive the parameters describing $\Gamma (z)$ implied by the evolution
in Lyman $\alpha$ forest line density.
We observe some flattening of $d{\cal N}/dz$ at $z < 1.7$, but
not to the degree seen by Weymann et al.\ (1998) in the Key Project data.
We find $\gamma$=0.50$\pm$0.21, for
lines above a 0.24 $\AA$ threshold (Paper IV), while Weymann et al.\ (1998) measure
$\gamma$=0.15$\pm$0.23.
We find $(A,B,z_{c},S)$ = (3.0$\times 10^{-12}$, 0.61, 5.5$\times 10^{-7}$, 7.07)
and (1.9$\times 10^{-11}$, 0.38, 3.4$\times 10^{-7}$, 6.21) for
($\Omega_{\rm M}$,$\Omega_{\Lambda}$)=(1.,0.) and
lines with rest equivalent widths above 0.24 and 0.32 $\AA$, respectively.
These fits to Equ.\ 6 are shown in Figure 3(a).
In panel (b), we plot $\Gamma (z)$, as expressed in Equ.\ 5, evaluated
using the parameters found from the fit to Equ.\ 6 above.
The HM96 solution and the
solution derived from the full FOS and MMT data sets are represented by the thick
and thin black lines respectively.
The small values of $z_{c}$ derived from $d{\cal N}/dz$ above translate into
ionization rates that do not decrease dramatically with decreasing redshift
and result from the less pronounced flattening of $d{\cal N}/dz$ relative
to the Key Project.
The observed $\Gamma (z)$ falls short of the ionization rate needed
to fully account for the change in the Lyman $\alpha$ line density with redshift,
indicating that the formation of structure in the low redshift universe
plays a significant role in determining the character of the Ly$\alpha$ forest
line density.

\section{Summary of Results}
\label{sec-summary}

We have analyzed a set of 151 QSOs and 906 Ly$\alpha$ absorption lines,
the subset of the total data set presented in Paper III that is
appropriate for the proximity effect.
The primary results of this work are as follows:

(1) The value of $J(\nu_{0})$ is observed to increase with redshift
over the redshift range of the sample data, $0.03 < z < 1.67$.
Dividing the sample at $z = 1$, we find $J(\nu_{0})$=
6.5$^{+38.}_{-1.6} \times 10^{-23}$ ergs s$^{-1}$ cm$^{-2}$ Hz$^{-1}$ sr$^{-1}$,
at low redshift and $J(\nu_{0})$=
1.0$^{+3.8}_{-0.2} \times 10^{-22}$ ergs s$^{-1}$ cm$^{-2}$ Hz$^{-1}$ sr$^{-1}$
at high redshift.

\begin{figure}
\plotfiddle{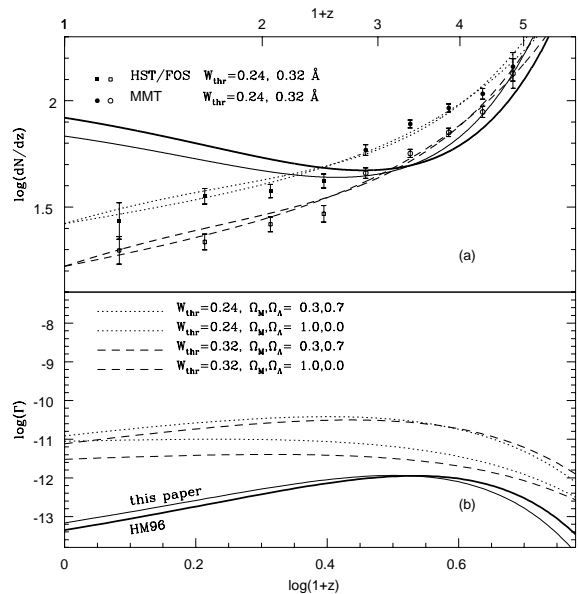}{2.75in}{0}{40}{40}{-130}{-75}
\caption{
(a) $d{\cal N}/dz$ versus redshift:
$W_{\rm thr}=$0.24 $\AA$ with fit to Equ.\ 6
(solid points, dotted lines),
$W_{\rm thr}=$0.32 $\AA$ with fit to Equ.\ 6
(open points, dashed lines),
Equ.\ 6 evaluated with HM96 parameters for
$\Gamma(z)$ expressed by Equ.\ 5 (thick solid line),
Equ.\ 6 evaluated with parameters for $\Gamma(z)$
found in this paper (thin solid line);
(b) $\Gamma(z)$ versus redshift expressed by
Equ.\ 5 using HM96 parameters (thick solid line),
using parameters found in this paper (thin solid line),
and using parameters found from
fits to $d{\cal N}/dz$ for
$W_{\rm thr}=$0.24 $\AA$ and ($\Omega_{M}$,$\Omega_{\Lambda}$)=(1.0,0.0) (thin dotted line),
$W_{\rm thr}=$0.24 $\AA$ and ($\Omega_{M}$,$\Omega_{\Lambda}$)=(0.3,0.7) (thick dotted  line),
$W_{\rm thr}=$0.32 $\AA$ and ($\Omega_{M}$,$\Omega_{\Lambda}$)=(1.0,0.0) (thin dashed line),
and
$W_{\rm thr}=$0.32 $\AA$ and ($\Omega_{M}$,$\Omega_{\Lambda}$)=(0.3,0.7) (thick dashed line)
\label{fig:dndz} }
\end{figure}

(2)  The inclusion of blazars at $z < 1$, damped Ly$\alpha$ absorbers, or
associated absorbers has no significant effect on the result.

(3) Using information measured and gathered from the literature on each QSO's UV spectral
index and solving for the HI ionization rate, yields
$1.9 \times 10^{-13}$ s$^{-1}$ for $z < 1$ and
$1.3 \times 10^{-12}$ s$^{-1}$ for and $z > 1$.
Solving directly for
the parameters $(A,B,z_{c},S)$ in the HM96 parametrization of $\Gamma(z)$
using the HST/FOS data presented in Papers III and IV, combined with the
high redshift, ground-based data presented in Papers I and II, results
in $(A,B,z_{c},S)$=($7.6\times 10^{-13}$,0.35,2.07,1.77) for $1.7 < z < 3.8$.

(4) The $z < 1$  result is in agreement with the range of values of
the mean intensity of the hydrogen-ionizing background allowed by
a variety of local estimates, including H$\alpha$ imaging and modeling of
galaxy HI disk truncations (Maloney 1993, Corbelli \& Salpeter 1993, Dove \& Shull 1994, 
Kutyrev \& Reynolds 1989, Tumlinson et al. 1999).  To within the uncertainty in the measurement,
this result agrees with the one previous proximity effect
measurement of the low redshift UV background (KF93).  These results are
consistent with
calculated models based upon the integrated emission from QSOs alone (HM96)
and with models which include both QSOs and starburst galaxies (Shull et al.\ 1999).
The uncertainties do not make a distinction between these two models possible.

(5) The results presented here tentatively confirm the IGM evolution scenario
provided by large scale hydrodynamic simulations (Dav\'{e} et al.\ 1999).  This
scenario, which is successful in describing many observed properties of the low redshift
IGM, is dependent upon an evolving $J(\nu_{0})$ which decreases from $z = 2$ to
$z = 0$.   However, the low redshift UV background required to match
the observations of the evolution of the Lyman $\alpha$ forest line density
is larger than found from the data, indicating that structure formation is
playing a role in this evolution as well.
Our results and the work of others are summarized in Figure 1.
We find some evidence of evolution in $J(\nu_{0})$, though it appears that even larger
data sets, especially at $z < 1$ and/or improved proximity effect ionization models
will be required to improve the significance.

\end{document}